\documentclass[twocolumn,prl,superscriptaddress,floatfix,aps,showpacs]{revtex4}
\usepackage{epsfig}
\usepackage{graphicx}
\usepackage{pgf}
\newcommand{\be}{\begin{equation}}
\newcommand{\ee}{\end{equation}}
\newcommand{\bea}{\begin{eqnarray}}
\newcommand{\eea}{\end{eqnarray}}
\newcommand{\bfr}{\mathbf r}

\newcommand{\bfQ}{\mathbf Q}
\newcommand{\bfk}{\mathbf k}
\newcommand{\bfs}{\mathbf s}

\begin{document}

\title{The Hartree-Fock ground state of the three-dimensional electron gas}
\author{Shiwei Zhang}
\affiliation{Department of Physics, College of William and Mary,
Williamsburg, VA 23187, USA}
\author{D.~M.~Ceperley }
\affiliation{NCSA and Department of Physics, University of Illinois at
Urbana-Champaign, Urbana, IL 61801, USA}

\begin{abstract}
  In 1962, Overhauser showed that within Hartree-Fock (HF) the electron gas is
  unstable to a spin density wave (SDW) instability.  Determining the true HF
  ground state has remained a challenge.  Using numerical calculations for
  finite systems and analytic techniques, we study the HF ground state of the
  3D electron gas. At high density, we find broken spin symmetry states with a
  nearly constant charge density.  Unlike previously discussed spin wave
  states, the observed wave vector of the SDW is smaller than $2 k_F$.  The
  broken-symmetry state originates from pairing instabilities at the Fermi
  surface, a model for which is proposed.
\end{abstract}

 \maketitle

The three-dimensional electron gas is one of the basic models of many-body
physics, and has been investigated for
over 70 years \cite{Wigner,Bloch,Overhauser,Trail,qmc-3deg,RMM_book,Giuliani}.
As the simplest model system representing an itinerant metal,
it consists of interacting electrons in a uniform 
neutralizing charge background, described by 
the Hamiltonian:
\be H = - {\hbar^2 \over 2m} \sum_i \nabla_i^2 + {1\over 2}\sum_{i\ne j}
{e^2 \over |\bfr_i-\bfr_j|} +
{\rm constant},
\label{eq:H_1stQ}
\ee
where the sums are over particle indices.
Its properties are routinely
used in density functional theory, e.g., in local density
approximations,
as a reference state in
calculations of electronic structure of real materials
\cite{RMM_book}. 

The simplest approach to an interacting many-fermion system such as
the electron gas (jellium) is the mean-field
Hartree-Fock (HF) method, 
which finds the Slater determinant wave function 
minimizing the variational energy.
In unpolarized jellium,
the ``conventional'' solution is a paramagnetic state
with spin symmetry,
the restricted HF (rHF)
solution in quantum chemistry.

The rHF solution, however, is not the exact HF ground state of jellium.  In
1962, Overhauser \cite{Overhauser} proved that the rHF solution is unstable
with respect to spin and charge fluctuations at any density.  The global
minimum energy state within HF is a spontaneously broken symmetry state.  The
properties of this global ground state have remained unknown \cite{Giuliani}.
This is surprising, given the fundamental importance of both the electron gas
and the HF approach.  The correlation energy of the homogeneous electron gas
is a commonly used fundamental concept, but its definition is in terms of the
HF energy of the electron gas.

In this paper, we numerically find the HF ground state
for finite systems.
Our motivation, aside from solving this 
mathematical puzzle,
was to understand the mechanism for the broken symmetry state.
Further, it was hoped that the solution would
suggest candidate
ground states, for jellium or for other systems,
that can then be explored by accurate many-body approaches such as
quantum Monte Carlo \cite{qmc-3deg,PW_afqmc}.
We focus on high and medium densities.
An analytic approach is used to augment and extend the results
to the thermodynamic limit.
We find that a different pairing instability characterizes the high-density 
ground state.

We consider $N$ ($N_\uparrow=N_\downarrow=N/2$) electrons in a cubic supercell
of volume $\Omega=L^3$.
The density is specified by the average distance between
electrons:
$r_s\equiv (3 \Omega/4\pi N)^{1/3}/a_B$.
We write a Slater determinant as
\be
|\Phi\rangle
=|\phi_1^\uparrow,\phi_2^\uparrow,\cdots,\phi_{N_\uparrow}^\uparrow\rangle
\otimes
|\phi_1^\downarrow,\phi_2^\downarrow,\cdots,
\phi_{N_\downarrow}^\downarrow\rangle,
\label{eq:det}
\ee
with 
$|\phi_j^\sigma\rangle=\sum_\bfk c_j^\sigma(\bfk)|\bfk\rangle$
where $|\bfk\rangle$ is a plane-wave basis function.
In the rHF solution,
any
$|\bfk\rangle$ with $k \equiv |\bfk| \le k_F$ 
is fully occupied, while all others
are empty. 
Our basis contains all plane-waves
with $k< k_{\rm cut}$.

To find the global ground state, i.e., the unrestricted HF (uHF) solution,
we use an
iterative projection
\be
|\Phi^{(m+1)}\rangle =
e^{-\tau H_{\rm HF}(\Phi^{(m)})} |\Phi^{(m)}\rangle,
\label{eq:HFproj}
\ee
where
$H_{\rm HF}(\Phi^{(m)})$ is the HF Hamiltonian, i.e.,
the mean-field approximation of Eq.~(\ref{eq:H_1stQ}).
The wave function remains a single Slater determinant in the
projection \cite{PW_afqmc}.
If $\tau$ is sufficiently small, the energy will decrease in each step
and the projection will converge to a HF solution
as $m\rightarrow \infty$.
To ensure
that the solution is not a local minimum,
we often start from multiple {\em random\/} initial states
 $|\Phi^{(0)}\rangle$ and verify that
the same final state is reached.

The smallness of the energy scale relevant to the symmetry breaking 
at high density presents a difficulty;
finite-size effects can easily be larger than
differences in energy of different phases.
As a consequence, the stable structures
vary wildly with $N$. 
There are subtle commensuration
effects in both $\bfr$-space (Wigner crystal) and $\bfk$-space (spin waves).
For example, in open-shell systems there is always a uHF
solution, since a broken symmetry state can be formed in a partially filled
shell to lower the exchange energy, with no cost to the
kinetic or Hartree energy (see below). 
In closed-shell systems,
on the other hand, there seems to be a critical value of $r_s^c(N)$, below
which no uHF state exists for a given value of $N$ under 
periodic boundary condition (PBC).

To break the shell
structure, 
we impose twisted boundary conditions \cite{tabc} 
on the orbitals (and hence on the wavefunction): $\phi(\bfr+L{\hat \alpha})
=e^{i2\pi\theta_\alpha}\phi(\bfr)$
where $\alpha=x,y,z$.
This applies a shift of ${\mathbf k_\theta}=2\pi{\vec \theta}/L$
to the plane-wave basis. 
${\vec \theta}=0$ corresponds to PBC, the $\Gamma$-point for solids.
For generic twist angles ${\vec \theta}$, 
the rHF solution is non-degenerate.

Let us write the Hamiltonian in second quantized form,
omitting an overall constant: 
\be
\hat H = {\hbar^2\over 2m} \sum_{\sigma,{\mathbf k}}
{\mathbf k}^2 c_{{\mathbf k},\sigma}^\dagger c_{{\mathbf k},\sigma}
+ {1\over 2\Omega}
\sum_{\Lambda}\,'
{4\pi e^2\over {\mathbf Q}^2}\,{\hat V}(\Lambda),
\label{eq:H_2ndQ}
\ee
where $c_{\bfk,\sigma}^\dagger$ and $c_{\bfk,\sigma}$ are creation and
annihilation operators. 
In the second term, 
$\Lambda$ denotes the variables 
$\{\bfk,\bfk',{\mathbf Q},\sigma,\sigma'\}$,
${\mathbf Q}$ 
is a reciprocal lattice vector,
the
$'$ on the summation indicates ${\mathbf Q}\ne0$, and
\be
\hat V (\Lambda) \equiv
c_{{\mathbf k-Q},\sigma}^\dagger c_{{\mathbf k'+Q},\sigma'}^\dagger
c_{{\mathbf k'},\sigma'} c_{{\mathbf k},\sigma}.
\label{eq:V2bdef}
\ee

The HF Hamiltonian
$\hat H_{\rm HF}(\Phi^{(m)})$
needed in Eq.~(\ref{eq:HFproj})
is the same as $\hat H$, 
but with $\hat V(\Lambda)$
replaced by the linearized form
$\hat V_{\rm HF}(\Lambda)=\hat v(\Lambda)-\langle \hat v(\Lambda)\rangle/2$,
where
\bea
\hat v(\Lambda) \equiv &
2\,\big[
\,\langle\sum_{\bfk''}
c_{\bfk'',\sigma'}^\dagger c_{\bfk''-\bfQ,\sigma'}\rangle
\delta_{\bfk',\bfk+\bfQ} \\ \nonumber
&-
\langle c_{\bfk'+\bfQ,\sigma'}^\dagger c_{\bfk+\bfQ,\sigma'}\rangle
\delta_{\sigma,\sigma'}
\,\big]\,
c_{\bfk,\sigma}^\dagger c_{\bfk',\sigma}.
\label{eq:VHFdef}
\eea
The expectation
$\langle..\rangle$ is with respect to 
$|\Phi^{(m)}\rangle$.
The variational energy is
$E_v(\Phi)\equiv \langle\Phi|\hat H|\Phi\rangle
/ \langle\Phi|\Phi\rangle$. 
We also compute a ``growth estimator'' of the energy,
$E_g\equiv - \ln [\langle\Phi^{(m+1)}|\Phi^{(m+1)}\rangle/
\langle\Phi^{(m)}|\Phi^{(m)}\rangle]/2\tau$,
in the projection.
At convergence, $E_v(\Phi^{(m)})=E_v(\Phi^{(m+1)})=E_g$, which means that
 $|\Phi^{(m)}\rangle$ is an eigenfunction of
$\hat H_{\rm HF}$. 
Hence, the projection gives a true solution of the
HF Hamiltonian, not just a Slater determinant with a variational
energy lower than the rHF value.

\begin{figure}
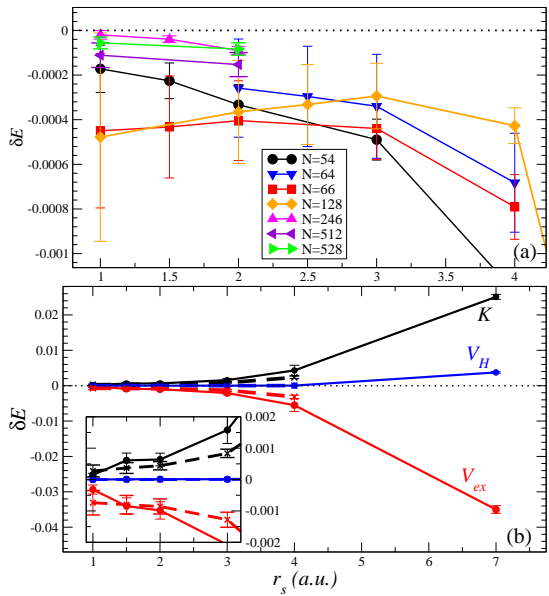

\begin{center}
\epsfig{file=fig1a.eps,width=0.4\textwidth,angle=0}
\epsfig{file=fig1b.eps,width=0.4\textwidth,angle=0}
\caption{(color online) Energy differences (in Ry)
between the HF ground state
and the rHF state. 
(a) The energy lowering per electron, $\delta E=(E-E_{\rm rHF})/N$,
 vs.~$r_s$ for different values of N \cite{note_frozen_core}.
(b) The three components of $\delta E$, 
kinetic ($K$), Hartree ($V_H$), and exchange ($V_{ex}$),
for $N=54$ (solid lines) and $N=66$
(dashed lines).
The inset is an enlargement of the low $r_s$. Error bars are estimated from the results for various ${\mathbf k}_\theta$-points.
}
\label{fig:Egain}
\end{center}
\end{figure}

Calculations were carried out with a set of random
$\vec \theta$ 
and the results averaged and errors estimated.
For each $N$, a fixed
set of $\bfk_\theta$-points were
used at different values of $r_s$ to correlate the runs.
At larger $r_s$, less statistical accuracy
is needed and fewer
$\bfk_\theta$-points were used.
Typically the plane-wave basis cutoff
was set to $k_{\rm cut} \sim 2$-$3k_F$, i.e.,
a kinetic energy cutoff of $5$-$10\,E_F$.  The resulting basis set error
is negligible for all but the largest $r_s$ ($=7$).
Fast Fourier transform (FFT) techniques were used to speed up each step
in the projection \cite{PW_afqmc}, and the orbitals 
re-orthonormalized as necessary.

The top panel of Fig.~\ref{fig:Egain} shows the energy difference $\delta E$
between the uHF ground state and the rHF state.
As $r_s$ is reduced,
the magnitude of $\delta E$ 
decreases before becoming nearly flat at high densities ($r_s<3$) 
for each value of $N$.
The energy lowering remains finite
across $r_s$,
showing that the broken symmetry uHF solution exists
for {\em all\/} densities,
consistent with Overhauser's proof \cite{Overhauser}.
At low $r_s$, $\delta E$ is roughly $-10^{-4}\,$Ry in the large-sized
systems. Calculations for larger systems are needed to clarify 
its behavior at small $r_s$. The energy differences are very small
compared to the rHF energy which, at $N\rightarrow \infty$, is $E_{\rm rHF}/N
=(2.21/r_s^2-0.916/r_s)\,$Ry: 
the {\it relative}
energy reduction vanishes as $r_s \rightarrow 0$.
From the bottom panel, 
we see that the lower total
energy of the uHF ground state
is achieved by reducing the
exchange energy at the cost of increased kinetic energy 
(exciting electrons into $|\bfk\rangle$ states with
$k>k_F$).
The Hartree energy remains unchanged up to $r_s\sim 3$.

\begin{figure}
\begin{center}
\epsfig{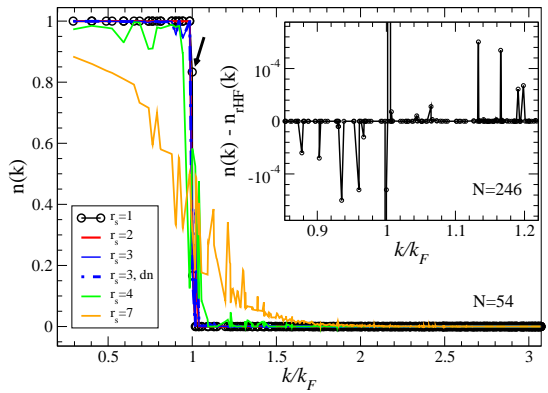}
\caption{(color online) Momentum distribution $n(\bfk)$
at different $r_s$ vs.~$k/k_F$.
The main graph has
$N=54$, with 
$\vec \theta=(-0.368,  0.172,  -0.364)$.
The arrow indicates $k_F$.
The inset shows $[n(\bfk)-n_{\rm rHF}(\bfk)]$ at $r_s=1$
for $N=246$, with $\vec \theta=(-0.494,  -0.425,  0.144)$.
Note the primary spike at $k_F$, and the  paired smaller
spikes with one on each side of $k_F$.
}
\label{fig:nG}
\end{center}
\end{figure}

We find that the momentum distribution in the uHF solution is
spin-independent, i.e.,
$n_\uparrow(\bfk)=n_\downarrow(\bfk)$
is an unbroken symmetry in the broken symmetry ground state.
This is illustrated in Fig.~\ref{fig:nG} for $r_s=3$,
but holds in all
our calculations when fully converged, at all $r_s$ and $N$.

Figure~\ref{fig:nG} also shows that,
as $r_s$ decreases,
the occupancy of $k>k_F$ states becomes
less pronounced, 
and significant modifications to the
Fermi sphere become increasingly
confined to the immediate vicinity of the Fermi surface (FS).
At small $r_s$, such modifications
tend to be principally single pairing states. An example is seen
at $r_s=1$ in $N=54$:
two plane-wave vectors
are involved,
such that
a pair of rHF orbitals
$|\phi^\uparrow\rangle=|\phi^\downarrow\rangle=|\bfk\rangle$ become
\begin{equation}
|\phi^\updownarrow\rangle    
=  c_\bfk |\bfk\rangle  \pm c_{\bfk'} |\bfk'\rangle
\label{eq:pairing}
\end{equation}
where $k\le k_F$ and $k'>k_F$, and $|c_\bfk|^2+|c_{\bfk'}|^2=1$.
Such a pairing state by itself forms a linear spin-density wave (SDW),
with constant charge density and
unchanged Hartree energy. As the inset
in Fig.~\ref{fig:nG} shows, additional pairing states
form in the uHF solution involving wave vectors further from the
FS. 
Although their amplitudes become very small as $r_s$
is reduced, these are important to the true uHF ground state,
as we discuss later.

\begin{figure}
\begin{center}
\epsfig{file=fig3a.eps,width=0.32\textwidth,angle=0}
\epsfig{file=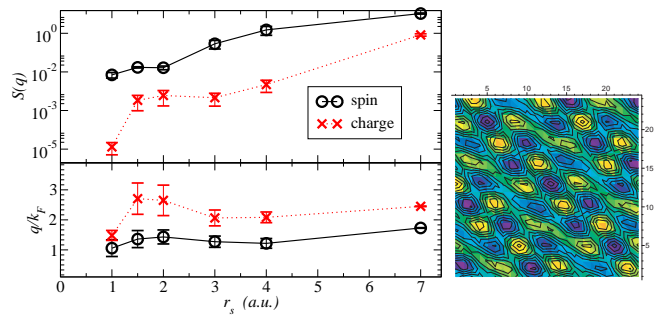,width=0.15\textwidth,angle=0}
\caption{(color online)
Left: 
Peak values and locations of the
Fourier transforms of spin and charge densities
for different values of~$r_s$ in $N=54$.
Errors are estimated from the values at different
$\bfk_\theta$-points. Note the logarithmic scale in $S(q)$.
Right: Contour plot of the spin density 
$\sigma(x,y,z)$
for 
a slice parallel to the $x$-$y$
plane. The system has 
$N=512$ and $r_s=1$.
The ${\mathbf q}$-vector 
$[2\bar{6}\bar{3}]$ has the leading  $S_\sigma({\mathbf q})$ value,
followed by $[33\bar{3}]$, and then $[4\bar{1}5]$ with $S_\sigma$ 
four times smaller.
}
\label{fig:nR}
\end{center}
\end{figure}

Real-space properties
are examined in Fig.~\ref{fig:nR}.
The charge density is
$\rho(\bfr)=n_\uparrow({\mathbf r})+ n_\downarrow({\mathbf r})$
and the spin density is
$\sigma(\bfr)=n_\uparrow({\mathbf r})- n_\downarrow({\mathbf r})$.
We measure their Fourier transforms, e.g.,
$S_\rho({\mathbf q})=|\int \rho(\bfr) e^{i{\mathbf q}\cdot\bfr}
d\bfr|^2/N$.
At $r_s=7$, the $N=54$
system is an antiferromagnetic bcc Wigner
crystal.
 As $r_s$ decreases, electrons become less localized and
less particle-like.
Fluctuation in the charge density becomes much smaller,
as indicated by the rapid reduction in $S_\rho({\mathbf q})$.
This is consistent
with the vanishing Hartree energy in Fig.~\ref{fig:Egain}.
Although $S_\sigma({\mathbf q})$ also decreases with $r_s$,
it is much larger and spin symmetry remains broken.
At $r_s=4$, the SDW no longer has a bcc structure, and its
symmetry between $x$, $y$, and $z$ is broken.
At small $r_s$, the electrons are highly delocalized
and wave-like.
Charge variations are effectively compensated for
by spatial
``double occupancy'' of $\uparrow$ and $\downarrow$ electrons,
as in the pairing state discussed in Fig.~\ref{fig:nG}.

We measure the characteristic wave vector of the spin or charge density wave
by
$q=|{\mathbf q}|$, where ${\mathbf q}$ is the peak position of
$S({\mathbf q})$.
A wave vector of
$q_\sigma\sim 2 k_F$ seems to
have always been assumed in previous investigations
of the SDW states \cite{Overhauser,Giuliani}. However, we find the maximum
spin ordering is at smaller wave vectors, as shown in Fig.~\ref{fig:nR}
for $N=54$.
Consistent results are seen for larger $N$, e.g. at $r_s=2$,
$q_\sigma/k_F=1.2(2)$, $1.4(2)$, $1.5(3)$, and $1.5(2)$
for $N=66$, $128$, $246$, and $528$,
respectively.

\begin{figure}
\begin{center}
\epsfig{file=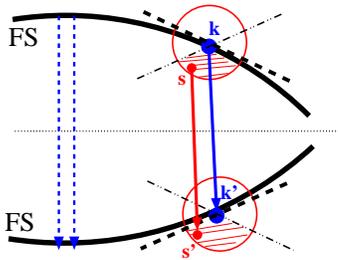,width=0.25\textwidth,angle=0}
\caption{(color online) Cartoon of pairing state with $q_\sigma<2k_F$.
The primary pairing state $\{\bfk,\bfk'\}$ is at the FS. The satellite
pairing state
$\{\bfs,\bfs'\}$ can be in the shaded areas, within $\mathcal{O}(2\pi/L)$ of
$\bfk$ and $\bfk'$. Dashed arrow lines illustrate
Overhauser\cite{Overhauser,Giuliani}
pairing at $2k_F$.
}
\label{fig:illus}
\end{center}
\end{figure}

We next prove analytically that an SDW instability
whose wave vector is $q_\sigma < 2k_F$ indeed exists.
Let us consider
a system of large but finite $N$, with $\bfk_\theta=0$.
 From the rHF reference state,
we create a broken symmetry state with two pairing orbitals
as in Eq.~(\ref{eq:pairing}),
using two points
at the FS, $\bfk$ (a highest occupied state) and
$\bfk'$ (a lowest unoccupied state),
as illustrated in Fig.~\ref{fig:illus}. The energy cost
consists of kinetic and exchange terms \cite{Giuliani}:
\begin{eqnarray}
\Delta K  & \sim &2 |c_{\bfk'}|^2 {\hbar^2 \over m} k_F\,\Delta k \\
\Delta V_{\rm ex}  & \sim & |c_{\bfk'}|^2
{e^2 \over \pi} \Delta k 
\ln{2k_F \over \Delta k},
\label{eq:pairing_dE}
\end{eqnarray}
where
$k_F=1/(\alpha r_s)$, with $\alpha=(4/9\pi)^{1/3}$.
Our choice of $\bfk$ and $\bfk'$ gives:
 $\Delta k=|\bfk'|-|\bfk| \sim (1/2l_F)(2\pi/L) = 1/(2l_F^2\alpha r_s)$,
where $l_F$ is defined by 
$k_F\equiv l_F\,(2\pi/L)$, i.e.,
$l_F=(3/8\pi)^{1/3}N^{1/3}$.
For fixed $r_s$,
the exchange term dominates if $N$ is sufficiently large.
Choosing
$c_{\bfk}$ and $c_{\bfk'}$ to be real and of $\mathcal{O}(1)$,
we can write Eq.~(\ref{eq:pairing_dE}) as:
\begin{equation}
\Delta V^{\bfk\bfk'}_{\rm ex} \sim
{e^2\over \pi \alpha r_s} {\ln l_F \over l_F^2}.
\label{eq:Vexkkp}
\end{equation}

We now create a ``satellite''
pairing state in the vicinity of $\bfk$ and $\bfk'$, i.e., with
$\bfk'-\bfk=\bfs'-\bfs$ and $|\bfs-\bfk|\sim \mathcal{O}(2\pi/L)$.
We choose the excitation amplitude
to be of the particular form:
$c_{\bfs'}\sim 1/(\ln l_F)^{1+\delta}$ ($\delta>0$), and thereby
$c_\bfs\sim 1$.
Using Eq.~(\ref{eq:pairing_dE}) and noting that
the difference in the magnitude of the wave vectors,
$\Delta s\equiv |\bfs'|-|\bfs|$, is less than
the size of circles in Fig.~\ref{fig:illus},
we obtain an upper bound to the energy cost
for creating the $\{\bfs,\bfs'\}$-pairing state
\begin{equation}
\Delta V^{\bfs\bfs'}_{\rm ex}
\sim {2 e^2\over \pi \alpha r_s} {1\over l_F\,(\ln l_F)^{1+2\delta}}.
\label{eq:Vexssp}
\end{equation}
The decrease in exchange energy because of
``constructive interference''
between the two parallel pairs
is:
\begin{eqnarray}
\Delta V^{\bfk\bfs}_{\rm ex}
&\sim & - 2 c_\bfk c_{\bfk'} c_\bfs c_{\bfs'}
{4\pi e^2\over L^3} {1\over |\bfk-\bfs|^2}   
\nonumber
\\
&\sim &- {e^2\over 2 \pi^2 \alpha r_s} {1\over l_F\,(\ln l_F)^{1+\delta}}.
\label{eq:Vexks}
\end{eqnarray}
For sufficiently large $l_F$, i.e., a large enough system size,
$|\Delta V^{\bfk\bfs}_{\rm ex}|$
can always be made
larger than the energy costs in Eqs.~(\ref{eq:Vexkkp}) and (\ref{eq:Vexssp}).
Hence 
this is an SDW state with lower energy than the rHF state.

The wave vector of the constructed SDW is ${\mathbf q}_\sigma=\bfk'-\bfk$.
As Fig.~\ref{fig:illus} shows, $q_\sigma$ does not need to be $2k_F$.
The angle between $\bfk'$ and ${\mathbf q}_\sigma$,
$\theta$, can range from $0$ ($q_\sigma=2k_F$) to $\pi/2$ ($q_\sigma=0$).
As $\theta$ increases,
more $\{\bfk,\bfk'\}$ pairing states become available on the FS,
while
the number of possible satellite pairs, i.e.,
the volume of the shaded areas, decreases. 
The optimal choice would be 
in between.
In fact, as a crude estimate, the number of $\{\bfk,\bfk'\}$ pairs
is 
$\propto 2\pi k_F^2 \sin\theta$, and the
number of $\{\bfs,\bfs'\}$ pairs for each is $\propto (\pi-\theta)$.
Maximizing their product gives $q_\sigma\sim 1.52 k_F$,
which is consistent with our data.

In previous approaches \cite{Overhauser,Giuliani},
pairing is constructed from orbitals directly across the FS.
The energy lowering is driven by
interference between such pairs 
(dashed arrows in Fig.~\ref{fig:illus}),
 which in our model
belong to primary pairing states.
Our approach differs by including the satellite pairing states.
The interference between the
primary and satellite states is 
what makes a general $q_\sigma$ possible.
This model is supported by the exact numerical data
in Fig.~\ref{fig:nG}.
Clearly, the true ground state goes beyond this model: the
energy will be further lowered by having more $\{\bfs,\bfs'\}$
pairs and multiple
$\{\bfk,\bfk'\}$ states, etc. 

We have shown that the uHF states at intermediate and
high densities
have nearly constant charge density. They are wave-like,
and arise from pairing between states on the FS
separated by distance $q_\sigma$.
The momentum distribution is
spin-independent. 
Its deviation from the Fermi sea
is increasingly confined to the vicinity of the FS as $r_s\rightarrow 0$.
The 
SDW wave vector is $q_\sigma\sim1.5(2)k_F$,
and its structures are determined
primarily by the short-range exchange potential \cite{note_nacl}.
In the rHF solution, the size of the exchange hole
is $r_x \equiv 2.34r_s$ \cite{RMM_book}.
A linear SDW with a wave length 
(i.e., characteristic like-spin separation) 
of $r_x$ has 
wave vector $q=1.40k_F$.
As $r_s\rightarrow 0$, 
the numerical results are
sensitive to the detailed topology of the FS in the finite-size systems.
The outcome of the SDW structure can vary greatly, as it 
is a delicate balance to optimize 
pairing among a small number of plane-wave states 
at the FS that can participate. 
Our results indicate that, at high densities,
the HF ground state
tends to further break spatial symmetry and favor
one or two dimensions.
As illustrated in the right panel of Fig.~\ref{fig:nR}, for example,
multiple 'pockets' can coalesce into locally or even globally 
(e.g., stripe-like) connected 
structures, in contrast with the Wigner crystal
state where each pocket, corresponding to one electron, is fully localized. 

To conclude, we have determined
the true HF ground state for finite electron gas. 
Combining numerical and analytic
results, we have described the origin and characteristics
of the broken symmetry state at high density,
and the novel pairing mechanism that drives it.

This work was supported by NSF (DMR-0535529 and DMR-0404853) and ARO
(48752PH).  We are grateful to H.~Krakauer for help with the plane-wave
machinery.  We acknowledge useful discussions with H.~Krakauer and
R.~M.~Martin.


\begin{thebibliography}{21}
\bibitem{Wigner} E. P. Wigner, Phys. Rev. {\bf 46}, 1002, (1934);
Trans. Faraday Soc. {\bf 34} 678 (1938).
\bibitem{Bloch} F. Bloch, Z. Phys. {\bf 57}, 549 (1929).
\bibitem{Overhauser} A. W. Overhauser, Phys. Rev. Lett. {\bf 3},
414 (1959); Phys. Rev. {\bf 128} 1427 (1962).
\bibitem{Trail} J. R. Trail {\it et.~al.\/} 
Phys. Rev. B {\bf 68}, 045107 (2003).
\bibitem{qmc-3deg}D. Ceperley, Phys. Rev. B {\bf 18}, 3126 (1978);
D.~M.~Ceperley and B.~J.~Alder, Phys. Rev. Lett. {\bf 45}, 566
(1980); F.~H.~Zong  {\it et.~al.\/}, 
Phys. Rev. E {\bf 66},
036703 (2002).
\bibitem{RMM_book} R.~M.~Martin, ``Electronic Structure: basic theory and
  practical methods'',
Cambridge University Press, 2004.
\bibitem{Giuliani} G. F. Giuliani and G. Vignale, ``Quantum Theory
of the Electon Liquid'', Cambridge University Press, 2005.
\bibitem{PW_afqmc} S. Zhang and H. Krakauer, Phys. Rev. Lett. {\bf 90},
  136401 (2003);
M.~Suewattana {\it et.~al.\/}, Phys.~Rev. B
{\bf 75}, 245123 (2007).
\bibitem{tabc}C.~Lin {\it et.~al.\/}, 
Phys.~Rev.~E {\bf 64}, 016702 (2001).
\bibitem{note_frozen_core}
Systems with up to $N=66$ were projected to convergence
following this approach.
Some larger systems were quenched from a ``frozen core'' state,
where $N_{\rm fc}$ ($< N$) electrons are frozen in the rHF state while the
remaining $N-N_{\rm fc}$ electrons are active in the projection.
$N_{\rm fc}$
is then gradually reduced in subsequent projections.
This procedure can sometimes get ``stuck'' in
a local minimum. It is thus possible that the energy gain in the larger systems
is underestimated and is a lower bound.

\bibitem{note_nacl}The structure which maximizes the near-neighbor
distance (NND) between 
like-spin electrons is
the close-packing fcc, which would lead to an NaCl-like SDW, but with 
a characteristic NND of
$2.28r_s$, less than $r_x$.

\end{thebibliography}
\end{document}